\documentclass{sf2a-conf2016}
\usepackage{graphicx}
\usepackage{hyperref}
\usepackage[]{natbib}  
\usepackage{epstopdf}

\def\BibTeX{{\rm B\kern-.05em{\sc i\kern-.025em b}\kern-.08em
    T\kern-.1667em\lower.7ex\hbox{E}\kern-.125emX}}
\bibpunct{(}{)}{;}{a}{}{,}  
\newcommand{\Lsol}{L$_{\odot}$}

\newcommand{\HI}{H\,{\sc {i}}~}

\newcommand{\Msold}{M$_{\odot}$\,yr$^{-1}$}

\newcommand{\kms}{km\,s$^{-1}$}

\begin{document}

\TitreGlobal{SF2A 2016}

\title{The axisymmetric envelopes of RS Cnc and EP Aqr}

\runningtitle{RS Cnc and EP Aqr}

\author{T. Le Bertre}\address{LERMA, UMR 8112, CNRS \& 
Observatoire de Paris/PSL, 61 av. de l'Observatoire, F-75014 Paris, France}

\author{D.\,T. Hoai$^{1,}$}\address{Department of Astrophysics, 
Vietnam National Satellite Center, VAST,  18 Hoang Quoc Viet, Ha Noi, Vietnam}

\author{P.\,T. Nhung$^{1,2}$}

\author{J.\,M. Winters}\address{IRAM, 300 rue de la Piscine, Domaine 
Universitaire, F-38406 St. Martin d'H\`eres, France}




\setcounter{page}{237}


\maketitle


\begin{abstract}
We report on observations obtained at IRAM on two semi-regular
variable Asymptotic Giant Branch (AGB) stars, RS Cnc and EP Aqr, 
undergoing mass loss at an intermediate rate of $\sim 10^{-7}$
\Msold. Interferometric data obtained with the Plateau-de-Bure 
interferometer (NOEMA) have been combined with On-The-Fly maps
obtained with the 30-m telescope in the CO(1-0) and (2-1) rotational 
lines. The spectral maps of spatially resolved sources reveal 
an axisymmetric morphology in which matter is flowing out at 
a low velocity ($\sim$ 2\,\kms) in the equatorial planes, and 
at a larger velocity ($\sim$ 8\,\kms) along the polar axes. There 
are indications that this kind of morpho-kinematics is relatively 
frequent among stars at the beginning of their evolution on 
the Thermally-Pulsing AGB, in particular among those that show 
composite CO line profiles, and that it might be caused by the
presence of a companion. We discuss the progress that could be 
expected for our understanding of the mass loss mechanisms in 
this kind of sources by increasing the spatial resolution of 
the observations with ALMA or NOEMA. 
\end{abstract}



\section{Introduction}
Stars on the Asymtotic Giant Branch (AGB) are in a short phase 
of their life (from 1 to a few 10$^6$ years). They evolve rapidly owing
to their large luminosity (few 10$^3$ \Lsol), and to the ejection of
their stellar envelopes. The mechanism by which stars expel their
envelopes is essential to the understanding of the terminal phases 
of stellar evolution, and to a proper description of their
contribution to the replenishment of the interstellar medium. 

Carbon monoxide (CO) is one of the best tracers of the winds from AGB
stars. It originates in the stellar atmospheres and survives up to a few
10$^{16-17}$\,cm where it is distroyed by UV photons from the interstellar
radiation field \citep{mamon1988}. It thus can be used to probe the 
region where the winds are shaped and accelerated. The first rotational
lines of CO (1-0 and 2-1) are easily accessible from the
Plateau-de-Bure (NOEMA), and higher degree lines can be observed from the
Atacama desert (ALMA). Modelling of the CO line profiles has provided 
the best mass loss rate estimates \citep[e.g. ][]{schoier2001,teyssier2006}. 

In the process of an investigation on the mass loss mechanisms 
\citep{winters2000,winters2003}, we became interested in sources 
that exhibit composite CO line
profiles, with a narrow component (FWHM $\sim$ 2-3 \kms) overimposed
on a broader  (FWHM $\sim$ 8-10 \kms) one. 
These sources were first pointed out by \citet{knapp1998} 
who suggested that the peculiar line profiles reveal two
different successive winds. We report on our recent work on 
two such cases, EP Aqr and RS Cnc. 
We selected these two stars for which a wealth of ancillary data is
available, and allows us to characterize their stages of evolution. 
Although our first investigation on EP Aqr \citep{winters2007} 
tended to support the Knapp et al. interpretation, 
our study of RS Cnc \citep{libert2010} showed that a composite CO
line profile might also result from an axi-symmetrical structure, with a
slow equatorial wind, and a rapid bipolar outflow. This encouraged us
to revisit both sources with a new modelling approach based on the
fitting of CO(1-0) and (2-1) spectral maps \citep{hoai2015}.  

\section{Observations}

\subsection{RS Cnc}
RS Cnc is an oxygen-rich S-type star (M6III), with an excess of s-process 
elements, a $^{12}$C/$^{13}$C abundance ratio of 35 
\citep{smith1986}, and Tc lines in its optical spectrum \citep{lebzelter1999}. 
It is clearly in the thermally pulsing phase of the AGB (TP-AGB) 
with dredge-up events. The Hipparcos parallaxe places it at a distance
of 143$_{-10}^{+12}$ pc \citep{vanleeuwen2007}. 
The associations with a far-infrared extended source, detected by IRAS 
and imaged by Spitzer \citep{geise2011}, 
and with an \HI emission detected with the Nan\c cay Radiotelescope
\citep{gerard2003} and imaged with the VLA \citep{matthews2007}, 
show that the central star has been undergoing mass loss for at least 
6$\times10^4$ years \citep{hoai2014}.

\subsection{EP Aqr}
EP Aqr is an oxygen-rich M-type star (M8III).
The Hipparcos parallaxe places it at a distance
of 114$_{-8}^{+8}$ pc \citep{vanleeuwen2007}. Its luminosity (3450 \Lsol)
and the low value of its $^{12}$C/$^{13}$C abundance ratio 
\citep[$\sim$ 10, ][]{cami2000} 
show that it is at the beginning of its evolution on the TP-AGB.  
Also, the absence of clear Tc lines in its spectrum \citep{lebzelter1999} 
suggests that it has not undergone any dredge-up event. 
It is associated with a far-infrared extended source, detected by IRAS 
and imaged by Herschel \citep{cox2012}, which shows that it has been
undergoing mass loss for more than a few 10$^4$ years. 

\subsection{Observational results and interpretation}
We have obtained spectral maps in the (1-0) and (2-1) rotational lines of CO 
by combining interferometric data from the Plateau-de-Bure Interferometer
with short spacing data from the Pico Veleta 30-m telescope. 
The data for RS Cnc have been presented by \citet{libert2010}  
and \citet{hoai2014}. Those for EP Aqr have been presented by
\citet{winters2007}, and reanalysed with a new processing  
by \citet{nhung2015a}. 

On the basis of characteristically shaped PV diagrams, the RS Cnc data
were interpreted by \citet{libert2010} as showing evidence 
for a bipolar geometry. Furthermore, 
the CO(1-0) channel maps around the central velocity (6.6 \kms),  
which have been obtained with the extended configurations (A and B) 
at high angular resolution, reveal a companion at 1$''$ north-west 
of the AGB star \citep{hoai2014}. As it is not seen in the continuum, 
it is presently not clear whether this companion is a compact 
(sub-)stellar object or a cloud in the circumstellar shell. 

The EP Aqr data exhibit a rather circular symmetry with enhancements 
of the emission at some distance from the central star that led
\citet{winters2007} to assume, as first suggested by \citet{knapp1998}, 
that the mass loss has been variable within the last few 10$^3$ years. 
However, the reanalysis by \citet{nhung2015a} showed that a morphology 
similar to that invoked for RS Cnc is also possible, and in fact
provides an even better account of the available data.

\section{Discussion}

The spatio-kinematic structure of RS Cnc has been reconstructed by
using a model of CO emission adapted to an arbitrary geometry 
\citep{hoai2014,hoai2015}. We use an axi-symmetric model in which 
the wind velocity and the flux of matter are smooth functions of 
the latitude, $\theta$. The wind is assumed to be stationary.
The excitation temperature is parametrized as a power of 
$r$, the distance to the central star. 

\begin{figure}[ht!]
 \centering
 \includegraphics[width=0.48\textwidth,clip]{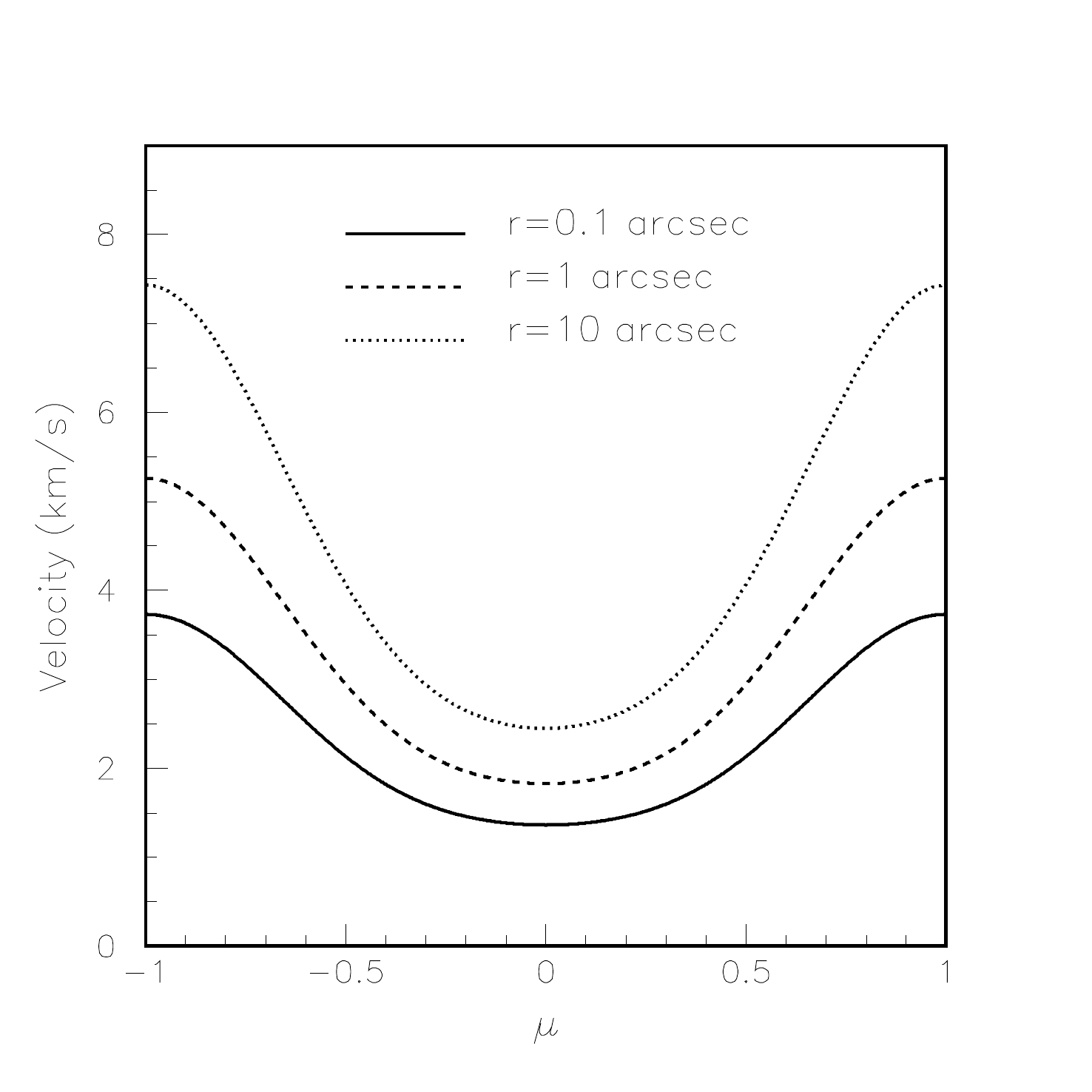}   
 \includegraphics[width=0.48\textwidth,clip]{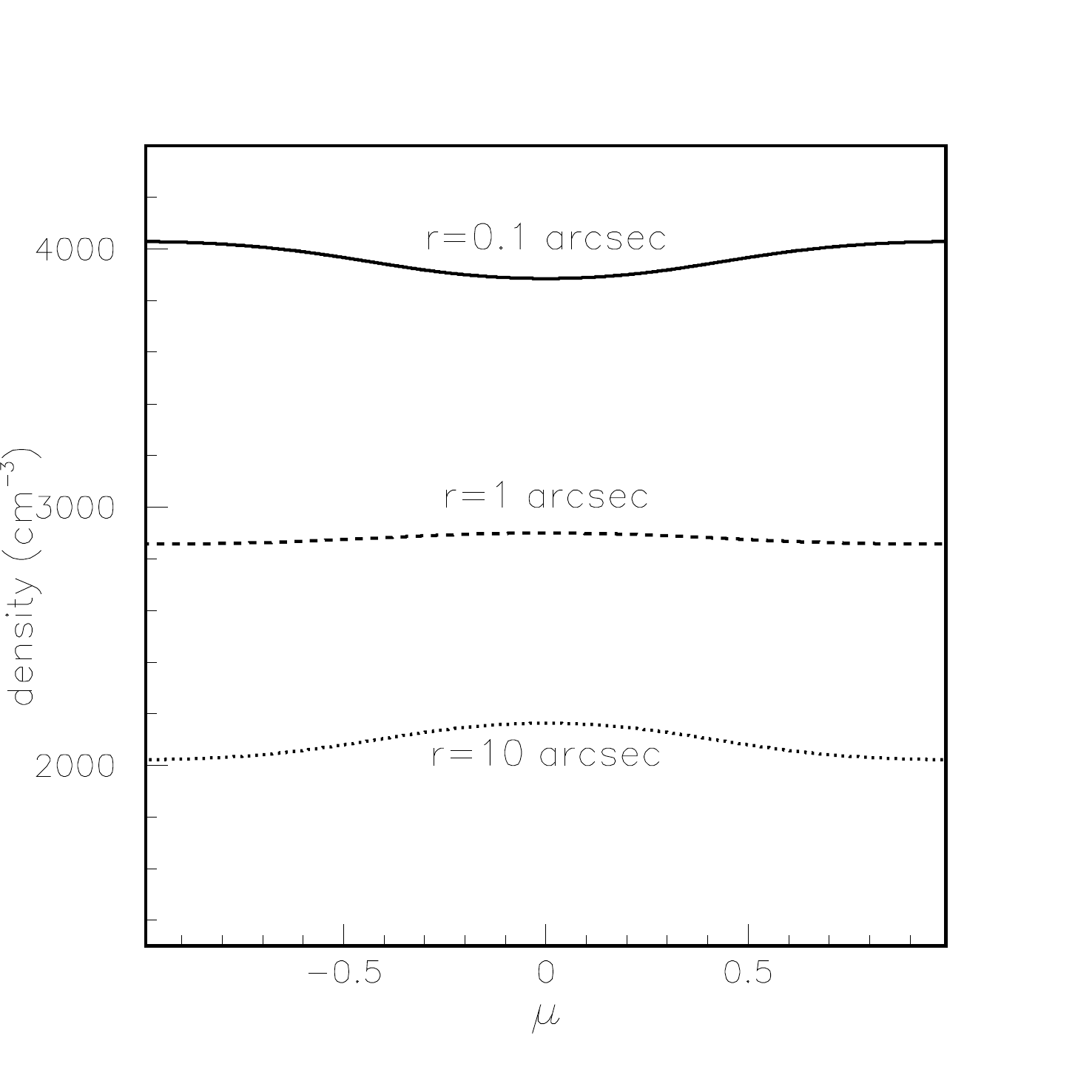}      
  \caption{RS Cnc morpho-kinematics \citep[from ][]{hoai2014}. 
{\bf Left:} velocity as a function of  
latitude ($\mu = cos \theta$). {\bf Right:}\,density (in number of
hydrogen atoms) as a function of latitude.}
  \label{lebertre:fig1}
\end{figure}

\begin{figure}[ht!]
 \centering
 \includegraphics[width=0.48\textwidth,clip]{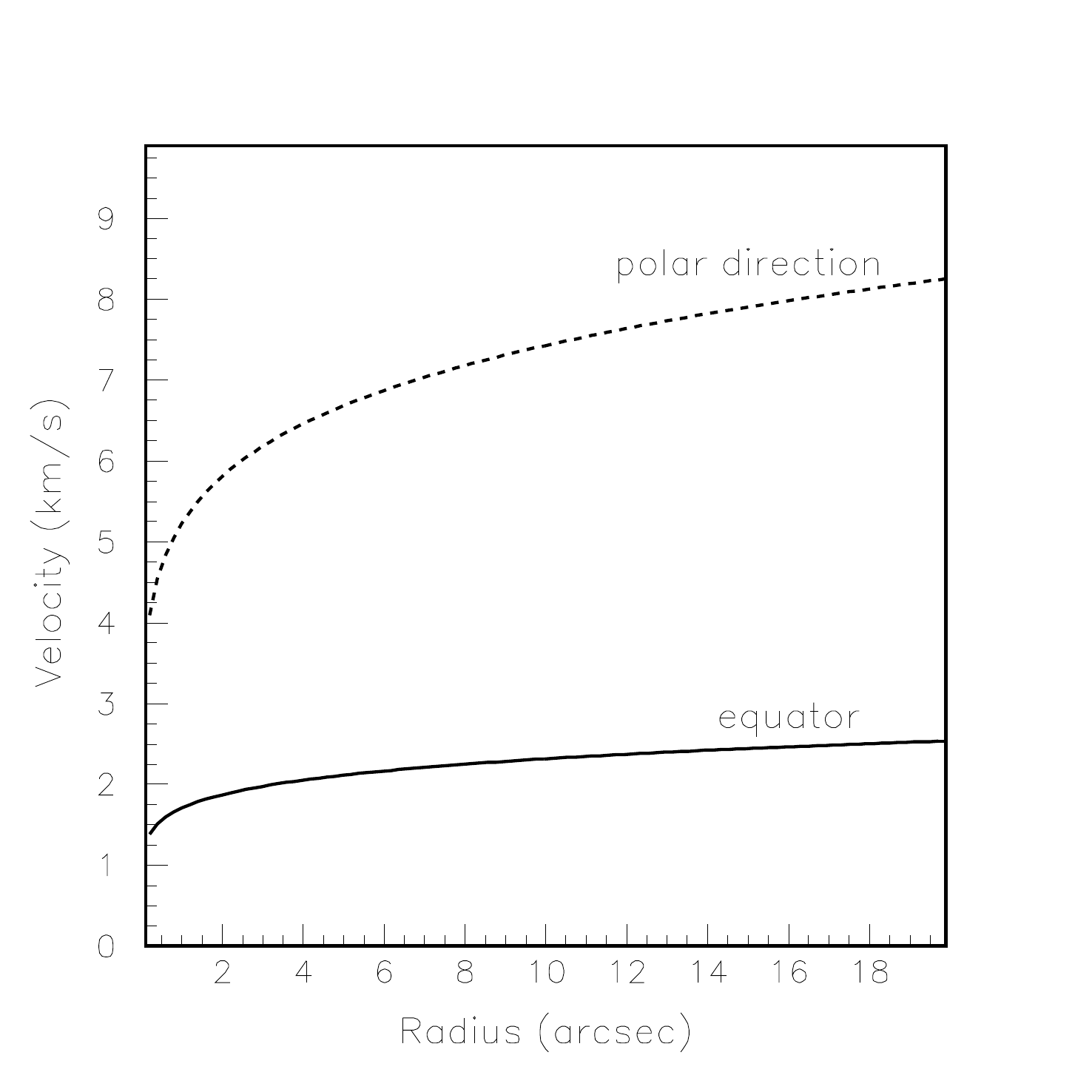}   
 \includegraphics[width=0.48\textwidth,clip]{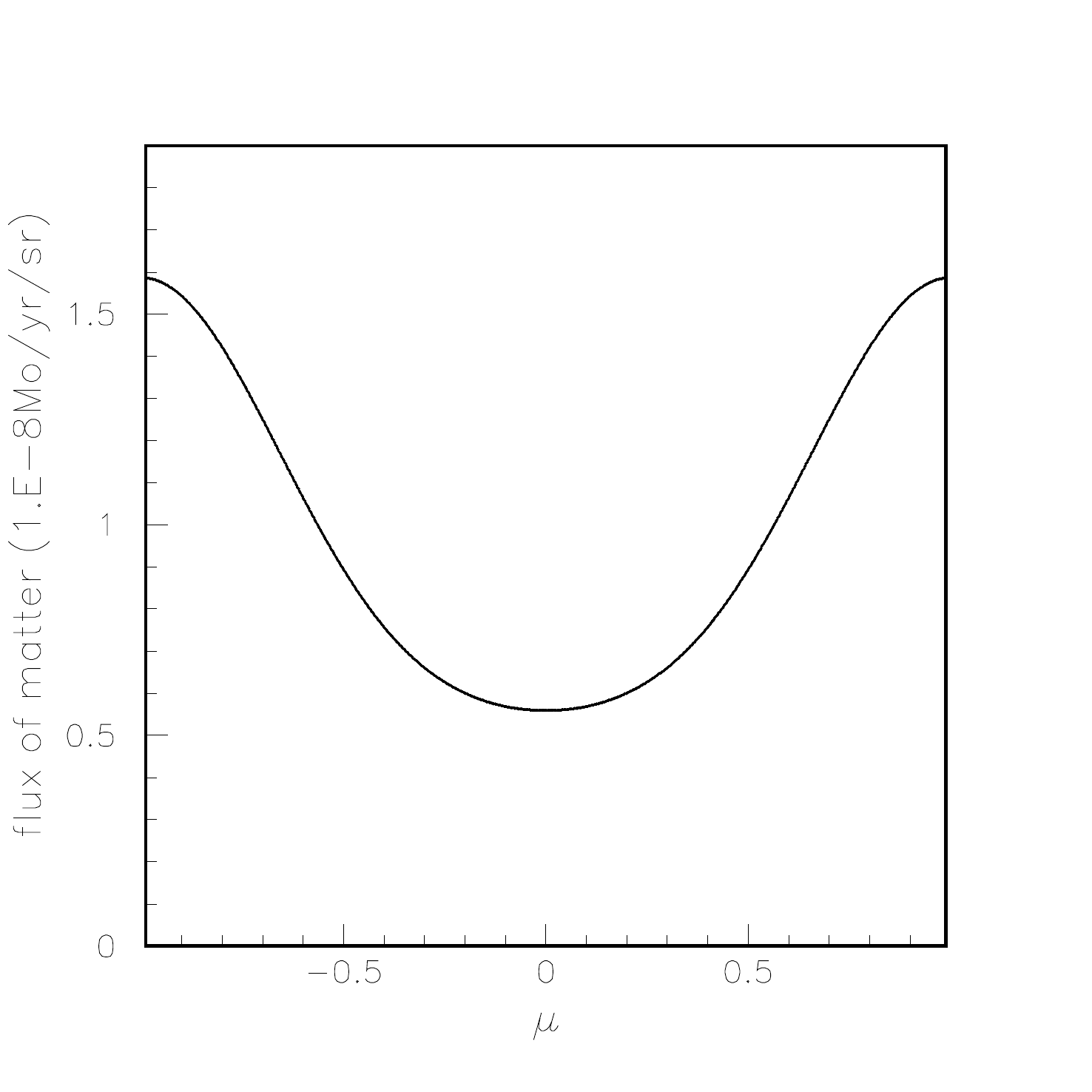}      
  \caption{RS Cnc \citep[from ][]{hoai2014}. 
{\bf Left:} velocity as a function of  
distance to the central star. {\bf Right:} flux of matter 
as a function of latitude ($\mu = cos \theta$).}
  \label{lebertre:fig2}
\end{figure}

The parameters of the model and its orientation in space are obtained 
by minimizing the sum of the square of the deviations (modelled minus 
observed intensities in the two spectral maps). We obtained a good fit
to the data with a model in which the velocity increases smoothly from 
the equatorial plane to the polar direction
(Fig.~\ref{lebertre:fig1}, left), whereas the density is almost independent 
of the latitude (Fig.~\ref{lebertre:fig1}, right). 
Another important implication of our study is that matter might 
still be accelerated at large distances (a few hundred AU, 
Fig.~\ref{lebertre:fig2}, left).

It is interesting to note that, in this model, the bipolarity is 
essentially apparent in the kinematics. It illustrates the importance 
of spectrally resolving emission lines. 

There are some slight deviations in the fits as compared to the
observations 
that can be further reduced by introducing a slight asymmetry in the 
polar flows of the model, with, south, a denser and faster flow 
than north \citep{nhung2015b}. 

The successful modelling of RS Cnc with an axi-symmetrical geometry called 
for a reconsideration of our initial interpretation of a variable mass loss 
for EP Aqr. Indeed, an axi-symmetrical geometry with a polar axis
pointing towards the observer could as well be considered for
accounting for the symmetry of the images projected on the plane 
of the sky. In \citet{nhung2015a} we studied this second option, 
and concluded that it is more likely than the first one. 
Interestingly, in the best model,  
the density does not deviate significantly from spherical symmetry, and, 
as for RS Cnc, the axi-symmetry shows up only in the velocity field. 

It is worth noting that the bipolarity which has been evidenced in 
RS Cnc and EP Aqr corresponds to an excess of flux of matter along the polar
directions (Fig.~\ref{lebertre:fig2}, right), and not to evacuated regions. 
Magnetic fields and binarity are commonly invoked to explain the
bipolarity observed in circumstellar shells.   
\citet{matt2010} have shown that a magnetic field should enhance the
mass loss towards the equatorial plane. However we observe the opposite 
(Fig.~\ref{lebertre:fig2}, right), and thus we tend to prefer the
second option (i.e. the effect of a companion). 

\citet{theuns1993} and \citet{mastrodemos1999} 
have studied the hydrodynamics of  stellar flows in binary systems. 
They find that for some configurations a three-dimensional shock 
wave propagates through the circumstellar shell and that it supports 
a spiral pattern. 
In the carbon star AFGL 3068, \citet{mauron2006} have found 
in dust-scattered galactic light a spiral pattern 
that fits well with the predictions of \citet{mastrodemos1999}. 
A spiral structure has also been detected in the CO(3-2) line emission
of R Scl by ALMA \citep{maercker2015}. The model  
predictions agree with a nearly spherical wind structure in the equatorial
plane on which the spiral pattern is imprinted. There are indications
in the EP Aqr channel maps of arcs, that may originate through such 
a mechanism. 

\citet{hoai2015} has also applied the RS Cnc modelling approach to X Her, 
another AGB star that shows composite CO line-profiles \citep{knapp1998}. 
The same kind of model fits satisfactorily well the data obtained at IRAM by
\citet{castro2010} on this source.

\section{Prospects}

To understand better the flow of matter from RS Cnc, we need to investigate 
the nature of the component (cloud\,?, companion\,?) located north-west
of RS Cnc. Broad-band observations would help to identify 
the presence of a compact stellar object
We need also to investigate the possibility of the presence of a
rotating disk, such as those found in some post-AGB stars
\citep{castro2012}. However, presently, there is no evidence
of such a kinematic structure in the data. 

For EP Aqr, we need also imaging at high spatial resolution 
($\sim$0.2$''$, or better) to probe the acceleration region of the polar
flows, as well as the structures in the equatorial plane (spiral ?), 
that in principle we should detect from a priviledged almost-polar 
line of sight. 

Although they do not bring the highest spatial resolution, 
the low-J lines allow to probe matter at large distances from the 
central star. They are also less affected by deviations from local 
thermal equilibrium and, especially in the inner regions of 
circumstellar envelopes, by optical depth effects, than the high-J lines. 

\section{Concluding remarks}
Axisymmetry seems to be a common feature in stellar winds, even in the
early phases of the TP-AGB. For instance, for EP Aqr the likely absence of Tc
in the atmosphere suggests that dredge-up events are still not
operating. It shows that non-spherical shapes observed often in planetary 
nebulae may arise from phenomena that are already active during the AGB
phase, although with effects which are less dramatic. 

In sources like RS Cnc or EP Aqr, the axi-symmetry 
becomes apparent only if the kinematical information is available.
The two important conclusions for these objects are that (i) the flux of
matter is larger along the polar axis than in the equatorial plane, 
(ii) the stellar winds might still be accelerated at a few hundred AU. 

We stress the importance of the high spectral resolution for studying 
the relatively slow winds from AGB stars. Indeed, we need to resolve
spectrally the emission for detecting the axi-symmetry of the source, 
but also this axi-symmetry may appear only as a kinematic effect (i.e. 
not as a density effect).  

The origin of bipolarity still needs to be identified. The fact that
we find sources with a flux of matter larger along 
the polar axis than in the equatorial plane does not favor 
magnetic fields. 
The presence of a companion should produce sub-structures, such as
spirals that can be detected in CO lines. It may affect the rate of 
mass loss, and may play an important role in the ultimate phases of
evolution of its primary. The presence of a rotating disk may have 
a considerable meaning for the mass loss mechanism, that up to now has
not been well explored in the models. 

High spatial resolution imaging at high spectral resolution of the
central parts of these sources where the winds are launched and 
accelerated with ALMA and NOEMA is clearly essential. 

\begin{acknowledgements}
{\bf{Acknowledgements}}\\
Based on observations carried out with the IRAM Plateau de Bure 
Interferometer and the IRAM 30-m telescope. IRAM is supported 
by INSU/CNRS (France), MPG (Germany) and IGN (Spain). We thank the
PCMI and the ASA programmes of the CNRS for financial support. 
\end{acknowledgements}

\bibliographystyle{aa}  
\bibliography{sf2a-template} 

\end{document}